\documentclass{sigchi}
\pdfoutput=1

\toappear{Permission to make digital or hard copies of all or part of this work for personal or classroom use is granted without fee provided that copies are not made or distributed for profit or commercial advantage and that copies bear this notice and the full citation on the first page. Copyrights for components of this work owned by others than the author(s) must be honored. Abstracting with credit is permitted. To copy otherwise, or republish, to post on servers or to redistribute to lists, requires prior specific permission and/or a fee. Request permissions from Permissions@acm.org. \\
{\emph{CHI'16}}, May 07 -- 12, 2016, San Jose, CA, USA. \\
Copyright is held by the owner/author(s). Publication rights licensed to ACM. \\
ACM 978-1-4503-3362-7/16/05\$15.00 \\
DOI: http://dx.doi.org/10.1145/2858036.2858400}

\usepackage{balance}  
\usepackage{txfonts}
\pdfmapfile{+txfonts.map}

\usepackage{times}    
\usepackage{color}
\usepackage{textcomp}
\usepackage{ccicons}
\usepackage{todonotes}

\usepackage{graphicx}
\usepackage[pdftex]{hyperref}
\usepackage{booktabs}
\usepackage{multirow}
\usepackage{multicol}
\usepackage{balance}
\usepackage{enumitem}
\usepackage[super]{nth}

\usepackage{url}

\newcommand*\rot{\rotatebox{90}}

\makeatletter
\def\url@leostyle{%
  \@ifundefined{selectfont}{\def\UrlFont{\sf}}{\def\UrlFont{\small\bf\ttfamily}}}
\makeatother
\def\pprw{8.5in}
\def\pprh{11in}

\definecolor{linkColor}{RGB}{6,125,233}
\hypersetup{%
  pdftitle={SIGCHI Conference Proceedings Format},
  pdfauthor={LaTeX},
  pdfkeywords={SIGCHI, proceedings, archival format},
  bookmarksnumbered,
  pdfstartview={FitH},
  colorlinks,
  citecolor=black,
  filecolor=black,
  linkcolor=black,
  urlcolor=linkColor,
  breaklinks=true,
}

\begin{document}

\title{``We're on the Same Page'': A Usability Study of\\Secure Email Using Pairs of Novice Users}

\numberofauthors{1}
\author{
\alignauthor
Scott Ruoti\raisebox{7pt}{$\dagger$}\titlenote{Sandia National Laboratories is a multi-program laboratory managed and operated by Sandia Corporation, a wholly owned subsidiary of Lockheed Martin Corporation, for the U.S. Department of Energys National Nuclear Security Administration under contract DE-AC04-94AL85000},
Jeff Andersen\raisebox{7pt}{$\dagger$},
Scott Heidbrink\raisebox{7pt}{$\dagger$}\raisebox{9pt}{$\ast$},
Mark O'Neill\raisebox{7pt}{$\dagger$}\raisebox{9pt}{$\ast$},
\\Elham Vaziripour\raisebox{7pt}{$\dagger$},
Justin Wu\raisebox{7pt}{$\dagger$},
Daniel Zappala\raisebox{7pt}{$\dagger$},
Kent Seamons\raisebox{7pt}{$\dagger$}\\
	\affaddr{
		Brigham Young University\raisebox{7pt}{$\dagger$},
		Sandia National Laboratories\raisebox{7pt}{$\ast$}
	}\\
	\email{ruoti@isrl.byu.edu, \{zappala, seamons\} @ cs.byu.edu}
}

\maketitle

\begin{abstract}
  Secure email is increasingly being touted as usable by novice users, with a push for adoption based on
  recent concerns about government surveillance. To determine whether secure email is
  ready for grassroots adoption, we employ a laboratory user study that recruits pairs of novice
  users to install and use several of the latest systems to exchange secure messages. We present
  both quantitative and qualitative results from 25 pairs of novice users as they
  use Pwm, Tutanota, and Virtru.
  Participants report being more at ease with this type of study and better able to cope with mistakes
  since both participants are ``on the same page''. We find that users prefer integrated solutions
  over depot-based solutions, and that tutorials are important in helping first-time users.
  Hiding the details of how a secure  email system provides security can lead to a lack of trust in the system.
  Participants expressed a desire to use secure email, but few wanted to use it regularly and most
  were unsure of when they might use it.
\end{abstract}

\keywords{Usable Security; Secure Email; User Study; Paired Participants}

\category{H.1.2.}{Models and Principles}{User/Machine Systems}[human factors]
\category{H.5.2.}{Information Interfaces and Presentation (e.g. HCI)}{User Interfaces}[user-centered design]

\section{Introduction}

In recent years there has been an increase in the promotion of secure email,
with tools such as Tutanota~\cite{tutanota},
Virtru~\cite{virtru},
ProtonMail~\cite{protonmail}, StartMail~\cite{startmail},
Hushmail~\cite{hushmail} and others
being pitched for everyday use by novice users. This interest is
likely spurred by concern over government surveillance of email,
particularly when third-party services such as Gmail and Hotmail
store email in plaintext on their servers. The Electronic Frontier
Foundation has heavily promoted secure communication and has released a security scorecard
of secure messaging systems that includes several email tools~\cite{effscorecard}.

While TextSecure, Signal, WhatsApp, and other secure instant messaging platforms are becoming popular,
it is unclear whether efforts to encourage users to likewise switch to
secure email will succeed, given that usable, secure email is still an unsolved problem more than fifteen years after it was first formally studied~\cite{whitten1999why}.
Moreover, widespread use of secure email partly depends on whether it could be adopted in a grassroots fashion, where both parties of an email conversation are novice users. All prior laboratory usability studies of secure email bring one novice user at a time into the lab and have him or her communicate with a study coordinator using a secure email system. While this helps researchers understand how well a novice user can start using secure email when paired with an expert user, it does not shed light on whether a pair of novice users can start using the system independently.

In this work, we conducted the first two-person study of secure email, with 25 pairs of novice users brought into the lab and asked to exchange secure email between themselves.
We asked participants to bring a friend with them, ensuring the participants already knew each other, in the hope that participants would behave more naturally.
Participants then used several secure email variants without any specific training or instructions on how to use the systems other than what the systems themselves provided. The main difference between this type of study and a traditional single-user study are that the participants played different roles (initiating contact versus being introduced to secure email) and that they interacted with another novice user and not a study coordinator.

In our study, we tested three different secure email systems: Pwm, Tutanota, and Virtru.
Each of these systems represents a different philosophy related to the integration of secure email with existing email systems.
Pwm integrates secure email with users' existing Gmail accounts, allowing them to compose and receive secure email with a familiar interface.
In contrast, Tutanota is a secure email depot that requires users to log into Tutanota's website to interact with their secure messages.
Virtru is a hybrid of these two approaches, allowing users who install the Virtru plugin to use secure email that is integrated with Gmail but also allowing non-Virtru users to receive encrypted email through a depot-based system on Virtru's website.

Our results and participant comments lead to the following contributions:

\begin{enumerate}

\item \textbf{Using pairs of novice users for an email usability study has several benefits.}
Having participants play different roles allowed us to gather data about different types of first-use cases (i.e., sending a secure email first vs receiving a secure email first).
In addition, participants exhibited more natural behaviors and indicated that they felt ``more at ease'', that they and their friend were ``on the same page'' or at the same level
of technical inexpertise, and that they did not feel discomfort from being ``under the microscope''.

\item \textbf{Hiding the details of how a secure messaging system provides security can lead to a lack of trust in the system.} This phenomenon was first noted in some of our earlier work~\cite{ruoti2013confused}, but those results are affected by multiple confounding factors \cite{atwater2015leading}. This paper provides further evidence that when security details are hidden from users, users are less likely to trust the system.
For example, although Pwm and Virtru utilize the same authentication method, Pwm completes authentication without user interaction, causing several users to doubt Pwm's security.
Similarly, participants like that Tutanota requires that an email be encrypted with a password since this makes it clear that the message was protected, unlike other systems that manage and use keys behind the scenes.

\item \textbf{Participants prefer integrated solutions over depot-based solutions.}
While to some it may be intuitive that the participants would prefer to continue using their existing email accounts, a number of depot-based systems have appeared recently (e.g., Tutanota, ProtonMail, StartMail).
Our results demonstrate that most everyday users strongly dislike using separate websites such as secure email depots to read their email.

\item \textbf{Tutorials are very important for users of secure email.} When asked what they liked about Pwm and Virtru, participants often reported that it was the tutorials presented alongside these systems.
The efficacy of these tutorials is shown by the fact that while using Pwm and Virtru, participants were able to quickly complete the study task whereas while using Tutanota---which lacks a tutorial---participants took on average 72\% longer to complete the study tasks, often making mistakes as they did so.

\item \textbf{Participants want the ability to use secure email but are unsure about when they would use it.}
Three-quarters of the participants in our study indicated that they wanted to be able to encrypt their email, but only one-quarter indicated that they would want to do so frequently.
Furthermore, when asked to describe how they would use encrypted email in practice, most participants were unsure, giving only vague references to how secure email might be useful.
This demonstrates a need for future research to establish whether the true problem facing the adoption of secure email is usability or that day-to-day users have no regular need to send sensitive data via email.

\end{enumerate}

\section{Related Work}

Whitten and Tygar~\cite{whitten1999why} conducted the first formal user study of a secure email system (i.e., PGP 5), which uncovered serious usability issues with key management and users' understanding of the underlying public key cryptography. They found that a majority of users were unable to successfully send encrypted email in the context of a hypothetical political campaign scenario. The results of the study took the security community by surprise and helped shape modern usable security research.

Replications of the Whitten and Tygar study were done by both Garfinkel and Miller~\cite{garfinkel2005johnny} and Sheng et al.~\cite{sheng2006why}.
Garfinkel and Miller showed that automatic key management was more usable than the manual key management present in the original experiment.
However, the study revealed that the tool ``was a little too transparent'' regarding its integration with Outlook Express. As a result, some users failed to read the instructions associated with visual indicators.
Sheng at al. demonstrated that despite improvements made to PGP in the seven years since Whitten and Tygar's original publication, key management was still a challenge for users. Furthermore, they showed that in the new version of PGP, encryption and decryption had become so transparent that users were unsure if a message they received had actually been encrypted.

More recently, we conducted a series of user studies with Private WebMail (Pwm), a secure email prototype that tightly integrates with the Gmail web interface~\cite{ruoti2013confused}.
Even though results showed the system to be quite usable, we found that some users made mistakes and were hesitant to trust the system due to the transparency of its automatic encryption.
We later revised Pwm to address the issues brought up in this earlier usability study~\cite{ruoti2015helping}.

In a replication of our work with Pwm, Atwater et al. verified that participants responded positively to automatic key management~\cite{atwater2015leading}.
They created a mock-up of Mailvelope that automatically generates keys for users, shares the public key with a key server, and automatically
retrieves an email recipient's public keys as needed.
Unfortunately, the mock-up lacked a working key management system, instead relying on hard-coded keys for email recipients, and 
did not simulate the delay that occurs when a sender has to wait for the recipient to generate and publish their public key. 
This made the mock-up incompatible with our study of first-time users, and it calls into question whether their results regarding the high usability of automated PGP apply to the common scenario of first-time PGP users.

Two prior studies included user interviews regarding secure email. Renaud at al.~\cite{renaud2014doesn} explored user's understanding of how email works and proposed some reasons why secure email adoption is low. Gaw et al.~\cite{gaw2006secrecy} interviewed users at a political activist organization that use secure email and noted that adoption was driven by the organization deciding encryption was necessary (due to secrecy concerns), and having IT staff setup the software for users enabled them to be successful. Even with this support, there were users who did not intend to use the software regularly, due to usability concerns and social factors.

\section{Secure Email}
Two and a half decades after the invention of PGP, secure email still remains sparsely used.
While some businesses require the use of secure email by their employees, there is little use of secure email by the population at large.
While it is possible that secure email will eventually diffuse from the workplace, it may be more likely that if secure email is to flourish, it will do so because of grassroots adoption (i.e., if participants are able to discover secure email on their own and easily begin using it with their acquaintances).

To date, no secure email studies have tested the ability of two novice users to begin using secure email; instead, these studies have tested a single novice user interacting with an expert user (i.e., a study coordinator).
Both Whitten and Tygar's study \cite{whitten1999why} as well as Garfinkel and Miller's \cite{garfinkel2005johnny} study used a simulated political campaign, where the study participant was the only individual in the campaign who did not already know how to use PGP.
Similarly, studies in Sheng et al.~\cite{sheng2006why}, our prior work~\cite{ruoti2013confused}, and Atwater et al.~\cite{atwater2015leading} involved participants sending email to study coordinators, none of which were instructed to simulate a novice user.

Even if the study coordinators had attempted to simulate a novice user, there are difficulties with this approach.
First, study coordinators are unlikely to make mistakes while using the encryption software, which is atypical of a true novice.
Even if study coordinators make use of scripted mistakes, there is a strong risk that these mistakes might be seen as artificial by participants, thereby breaking immersion for the participant.
Second, in many tasks there is a high level of variability possible in participant actions, making it difficult to script for all possible situations, and unscripted responses from coordinators are likely to be biased by their experience with the system.
Third, participants are likely to attribute any problems they encounter to their own mistakes, and not to the coordinator, whereas when interacting with a friend, participants are just as likely to attribute the mistake to their friend as to themselves.\footnote{In some ongoing work, we have attempted to simulate a novice user and encountered these difficulties in practice~\cite{ruoti2015helping}}.


To avoid these difficulties, our study uses two novice participants.
This is the first study to test whether two novice participants, who know each other beforehand, can successfully use secure email in a grassroots fashion.
Our observations, as discussed later in this paper, show that this approach produces more natural behavior then when participants email a study coordinator.
Moreover, this approach allowed us to examine how users perform when they are introduced to secure email in different ways (i.e., installing and then sending an email vs. receiving an email and then installing).

To select which systems to test, we surveyed existing secure email systems, including those listed on the EFF's scorecard~\cite{effscorecard}, and filtered them according to two criteria.
First, we focused on browser-based solutions, as previous work has shown that this approach is preferred by users~\cite{ruoti2013confused,atwater2015leading}.
Second, we required the systems to use automatic key management, as research has shown that users are highly amenable to this approach~\cite{garfinkel2005johnny,ruoti2013confused}.\footnote{We also investigated systems based on PGP, but found that these systems all had low usability and chose not to include them in this study (see {\bf Pilot Study}).}
Of the systems that matched these criteria, we found that they could be grouped into three types of secure email systems: integrated, depot-based, and a hybrid of integrated and depot-based systems.
For each of these groups, we tested the systems in the group, and selected the system that we felt had the best usability and included that system in our study.
The remainder of this section describes the types of secure email that were tested, as well as the representative system for each.

\subsection{Integrated Secure Email (Pwm)}
``Integrated secure email'' refers to secure email systems that integrate with users' existing email systems.
In this model, users do not need to create new accounts and are able to encrypt messages within the email interfaces they are already accustomed to \cite{ruoti2013confused}.

Private WebMail (Pwm)\footnote{\url{https://pwm.byu.edu/}} is the representative system for this type of secure email.
Pwm was developed as part of our research~\cite{ruoti2013confused, ruoti2015helping} and has the highest usability\footnote{Based on the System Usability Scale~\cite{brooke1996sus}.} of any secure email system tested in the literature~\cite{ruoti2015helping}.
Similarly, because Pwm has been studied before, it provides a good baseline for comparing the results of the other systems tested in this study.

Pwm is a browser extension that tightly integrates with Gmail's web interface to provide secure email. Users are never exposed to any cryptographic operation, including the verification of the user's identity, which is completed without user interaction.
Pwm modifies the color scheme of Gmail for encrypted emails in order to help users identify which messages have been encrypted.
Pwm also includes inline tutorials that instruct users on how to operate Pwm.

Pwm's threat model is focused on protecting email from individuals who do not have access to the sender's or recipient's email account.
While this does not protect email against attackers who compromise the user's email account, it does provide security during transmission and storage of the email.
Pwm is susceptible to a malicious email service provider.

\subsection{Depot-Based Secure Email (Tutanota)}

``Depot-based secure email'' refers to secure email systems that use a separate website from users' existing email systems.
In this model, users have a separate account with the depot where they can send and receive secure emails.
When a user receives a new message in their depot account, many depot-based systems will send an email to the user's standard email address, informing them that they have a new email to check in the depot system.
Often, these systems do not allow users to send email to individuals not already using the depot.
Depot-based systems are commonly deployed by companies and organizations for secure communication.

While there are many depot-based systems to choose from, most are either costly (e.g., Hushmail and StartMail) or are currently not offering email addresses to new users (e.g., ProtonMail).
We chose Tutanota\footnote{\url{https://tutanota.com/}} because it was the most usable of the depot systems we tested, is free, is currently available to new users, and is receiving publicity on Twitter.

Tutanota assigns users an email address ending in ``@tutanota.com''.
Users can send and receive email from this address as they normally would.
During account creation, Tutanota generates a public/private key pair for the user.
These keys are stored on Tutanota's servers, with the private key being encrypted with the user's Tutanota account password.
When Tutanota users send messages to other Tutanota users, the messages are automatically encrypted and signed with the appropriate keys.
When a Tutanota user sends a message to a non-Tutanota user, they have the option of encrypting it with a shared secret (i.e., password).
When the non-Tutanota user receives the encrypted email, they are redirected to Tutanota's website, where they can enter the shared secret and decrypt the message.
Tutanota's interface also allows the non-Tutanota user to respond to the message, and will encrypt the message using the same shared secret.

The threat model for Tutanota is similar to Pwm, except that instead of having normal and secure email stored in the same email accounts, they are stored in separate accounts.
This means that if a user's normal email account is broken into, their sensitive messages are still secure.
Users are still susceptible to having their secure email account password guessed/stolen or to a malicious email service provider.

\subsection{Hybrid Secure Email (Virtru)}
Virtru\footnote{\url{https://www.virtru.com/}} is a hybrid of integrated and depot-based secure email.
Once Virtru's browser plugin is installed, it functions much the same as Pwm, including automatic key management and integration with Gmail.
If a Virtru user sends an email to a non-Virtru user, the sender still does so through Gmail, but the recipient will receive an email informing them that they need to log into Virtru's website to view their message.
At this point Virtru is similar to Tutanota in its management of new users, except that instead of providing a password, non-Virtru users are asked to prove that they own their email address.
As such, the threat model for Virtru is identical to Pwm.

\section{Methodology}
We conducted an IRB-approved user study wherein pairs of participants used secure email to communicate sensitive information to each other.
This section gives an overview of the study and describes the scenario, task, study questionnaire, and post-study interview. In addition, we discuss the development and limitations of the study.

\subsection{Study Setup}
The study ran for two weeks---beginning Tuesday, September 8, 2015 and ending Friday, September 18, 2015.
In total, 25 pairs of participants (50 total participants) completed the study.
Participants took between forty and sixty minutes to complete the study, and each participant was compensated \$15 USD for their participation.
Participants were required to be accompanied by a friend, who served as their counterpart for the study.
For standardization and requirements of the systems tested in the study, both participants were required to have Gmail accounts.

When participants arrived, they were read a brief introduction detailing the study and their rights as participants.
Participants were informed that they would be in separate rooms during the study and would use email to communicate with each other.\footnote{The study coordinators ensured that the participants knew each other's email addresses.}
Participants were also informed that a study coordinator would be with them at all times and could answer any questions they might have.

Using a coin flip, one participant was randomly assigned as Participant A (referred to as ``Johnny" throughout the paper) and the other as Participant B (referred to as ``Jane" throughout the paper).
The participants were then led to the appropriate room to begin the study; each room had identical equipment.
For the remainder of the study, all instructions were provided in written form.
Participants completed the task on a virtual machine (VM), which was restored to a common snapshot after each study task, ensuring that the computer started in the same state for all participants and that no participant information was accidentally stored.

During the study, participants were asked to complete a multi-stage task three times, once for each of the secure email systems being tested: Pwm, Tutanota, and Virtru.
The order in which the participants used the systems was randomized.
For each system, participants installed any necessary software and were then given fifteen minutes to complete the task.
If they were unable to complete the task in the time limit, the study coordinators helped them move to the next system.
In practice, this only occurred a single time.

\subsection{Demographics}
We recruited Gmail users for our study at a local university.
Participants were two-thirds female: female (33; 66\%), male (17; 34\%).
Participants skewed young: 18 to 24 years old (44; 88\%), 25 to 34 years old (6; 12\%).

We distributed posters across campus to avoid biasing our results to any particular major.
All participants were university students,\footnote{We did not require this.} with the majority being undergraduate students: undergraduate students (40; 80\%), graduate students (10; 20\%).
Participants were enrolled in a variety of majors, including both technical and non-technical majors.
No major was represented by more than four participants, with the vast majority only having one or two participants.

\subsection{Scenario Design}
During the study, participants were asked to role-play a scenario about completing taxes. Each participant was shown the following text, respectively.

\begin{itemize}

\item \textbf{Johnny.}
Your friend graduated in accounting and you have asked their help in preparing your taxes. They told you that they needed you to email them your last year's tax PIN and your social security number. Since this information is sensitive, you want to protect (encrypt) this information when you send it over email.

\item \textbf{Jane.}
You graduated in accounting and have agreed to help a friend prepare their taxes. You have asked them to email you their last year's tax PIN and their social security number.

\end{itemize}

Participants were provided with the information they would send (e.g., SSN and PIN), but were told to treat this information as they would their own sensitive information.

\subsection{Task Design}

Based on the scenario, participants were asked to complete a three-stage task.

\begin{enumerate}

\item Johnny would encrypt and send their SSN and last year's tax PIN to Jane.
\item Jane would reply to this sensitive information with a confirmation code and this year's tax PIN. This information would also be encrypted.
\item Johnny would reply and let Jane know he had received the confirmation code and last year's tax PIN.

\end{enumerate}

The instructions guiding the participants through the three stages are as follows:

\begin{itemize}

\item \textbf {Johnny.}
In this task, you'll be using \{Pwm, Virtru, or Mailvelope\}. The system can be found at the following website: \{Appropriate Website\}. 
Please encrypt and send the following information to your friend using \{Pwm, Virtru, or Mailvelope\}:
SSN: \{Generated SSN\}.
PIN: \{Generated PIN\}.

Once you have received the confirmation code and PIN from your friend, send an email to your friend letting them know you have received this information. After you have sent this confirmation email, let the study coordinator know you have finished this task.

\item \textbf {Jane -- Sheet 1.}
Please wait for your friend's email with their last year's tax PIN and SSN. Once you have written down your friend's SSN and PIN, let the study coordinator know that you are ready to reply to your friend with their confirmation code and PIN.

\item \textbf {Jane -- Sheet 2.}
You have completed your friend's taxes and need to send them the confirmation code and this year's tax PIN from their tax submission. Since your friend used \{Pwm, Virtru, or Tutanota\} to send sensitive information to you, please also use \{Pwm, Virtru, or Tutanota\} to send them the confirmation code and PIN.
Confirmation Code: \{Generated code\}.
PIN: \{Generated PIN\}.

Once you have sent the confirmation code and PIN to your friend, wait for them to reply to you and confirm they received the information. Once you have received this confirmation, let the study coordinator know you have finished this task.

\end{itemize}

The instructions for Johnny and Sheet 1 of the instructions for Jane were given at the start of the task.
Sheet 2 for Jane was given once Johnny had received and decrypted the sensitive information sent by Jane in Stage 1.
Participants completed this task once for each of the three systems being tested.
Each time, the instructions only included information relevant to the system being tested.

While participants waited for email from each other, they were told that they could browse the Internet, use their phones, or engage in other similar activities.
This was done to provide a more natural setting for the participants, and to avoid frustration if participants had to wait for an extended period of time while their friend figured out an encrypted email system.

Study coordinators were allowed to answer questions related to the study but were not allowed to provide instructions on how to use any of the systems being tested.
If participants became stuck and asked for help, they were told that they could take whatever steps they normally would to solve a similar problem. Additionally, when asked for help, if the study coordinator believed communication between the two parties could help, he could remind participants that they were free to communicate with their friend and that only the sensitive information was required to be transmitted over secure email.

\subsection{Study Questionnaire}
We administered our study using the Qualtrics web-based survey software.
Before beginning the survey, participants answered a set of demographic questions.
Participants then completed the study task for each of the three secure email systems. 

Immediately upon completing the study task for a given secure email system, participants were asked several questions related to their experience with that system.
First, participants completed the ten questions from the System Usability Scale (SUS)~\cite{brooke1996sus,from2013sus}.
Multiple studies have shown that SUS is a good indicator of perceived usability~\cite{tullis2004comparison}, is consistent across populations~\cite{ruoti2015authentication}, and has been used in the past to rate secure email systems~\cite{ruoti2013confused,atwater2015leading}.
After providing a SUS score, participants were asked to describe what they liked about each system, what they would change, and why they would change it.

After completing the task and questions for all three secure email systems, participants were asked to select which of the encrypted email systems they had used was their favorite, and
to describe why they liked this system.
Participants were next asked to rate the following statements using a five-point Likert-scale (Strongly Disagree---Strongly Agree): ``I want to be able to encrypt my email,'' and ``I would encrypt email frequently.''

\subsection{Post-Study Interview}
After completing the survey, participants were interviewed by their respective study coordinator.
The coordinator asked participants about their general impressions of the study and the secure email systems they had used.
Furthermore, the coordinators were instructed to note when the participants struggled or had other interesting events occur, and during the post-study interview the coordinators reviewed and further explored these events with the participants.

After the participants completed their individual post-study interviews, they were brought together for a final post-study interview.
First, participants were once again asked which system was their favorite and why.
This question was intended to observe how participants' preferences might change when they could discuss their favorite system with each other.
Second, participants were asked to describe their ideal secure email system.
While participants are not system designers, our experience has shown that participants often reveal preferences that otherwise remain unspoken.
Finally, participants were asked to share their opinions related to doing a study with a friend.
They were informed that it was the first time that we had conducted such a study.
This question was designed to learn possible benefits and limitations of conducting such a two-person study.

\subsection{Pilot Study}
We conducted a pilot study with three pairs of participants (six participants total).
The lessons learned during the pilot study motivated two minor changes to the study.
First, the pilot study included Mailvelope, a PGP-based secure email system, in addition to the other systems.
In the pilot study, participants rated Mailvelope as having low usability, and even with prior PGP experience participants took between fifteen and thirty minutes to complete the task.
For these reasons, it was clear that Mailvelope was not compatible with our study setup (which was limited to one hour), and we did not include it the final study.\footnote{Instead, we ran a separate study that used the same methodology, but only examined Mailvelope~\cite{ruoti2015johnny}}
Second, in the pilot study, participants were shown all instructions within the Qualtrics survey.
After the pilot, we printed out the task instructions and gave these to users for easier reference.

\subsection{Limitations}
During the study, a bug in the Qualtrics software led to an uneven distribution of treatments (i.e., order the systems were used).
Due to this problem, treatments where Virtru was the first system tested made up two-thirds (68\%, n=17) of the studies. 
Other than this abnormality, treatment distribution was as expected.
We examined all of our qualitative data and, after adjusting for this abnormality, found no statistically significant difference and only one observable difference, which we note in the Results section.

Our study also has limitations common to all existing secure email studies.
First, our populations are not representative of all groups, and future research could broaden the population (e.g., non-students, non-Gmail users).
Second, our study was a short-term study, and future research should look at these issues in a longer-term longitudinal study.
Third, our study is a lab study and has limitations common to all studies run in a trusted environment \cite{milgram1978obedience,sotirakopoulos2010did}.

%
%

\section{Results}

In this section, we report the quantitative results from our user study.
First, we report on the usability scores for each system.
Next, we give the time taken to complete the task for each system as well as the number of mistakes encountered while using each system.
Finally, we report which system participants indicated was their favorite.

The data from our study, along with screenshots of each system, is available at  \url{http://chi2016.isrl.byu.edu/}.

\subsection{System Usability Scale}

\begin{table}
\centering
\resizebox{\columnwidth}{!}{

\begin{tabular}{l|l|c|cc|ccccc|}

	\rule{0pt}{11ex} & \rot{Participant} & \rot{Count} & \rot{Mean} & \rot{\shortstack[1]{Standard\\Deviation}} & \rot{Min} & \rot{Q1} & \rot{Median} & \rot{Q3} & \rot{Max} \\
	\midrule
	
	Pwm			& Johnny 	& 25 & 76.3 & 15.3 & 35.0 & 72.5 & 75.0 & 82.5 & 100 \\ 
	Pwm			& Jane	 	& 25 & 69.1 & 17.2 & 30.0 & 32.5 & 70.0 & 80.0 & 100 \\ 
	Pwm			& Both 		& 50 & 72.7 & 16.5 & 30.0 & 65.0 & 72.5 & 82.5 & 100 \\ 
	\midrule	
	
	Virtru		& Johnny 	& 25 & 73.1 & 14.7 & 27.5 & 70.0 & 75.0 & 80.0 & 97.5 \\ 
	Virtru		& Jane	 	& 25 & 71.4 & 12.7 & 40.0 & 67.5 & 75.0 & 77.5 & 87.5 \\ 
	Virtru		& Both 		& 50 & 72.3 & 13.7 & 27.5 & 67.5 & 75.5 & 80.0 & 97.5 \\ 
	\midrule	
	
	Tutanota	& Johnny 	& 25 & 50.0 & 18.2 & 27.5 & 35.0 & 50.0 & 62.5 & 92.5 \\ 
	Tutanota	& Jane	 	& 25 & 54.3 & 17.4 & 22.5 & 42.5 & 55.0 & 35.0 & 90.0 \\ 
	Tutanota	& Both 		& 50 & 52.2 & 17.8 & 22.5 & 40.0 & 25.5 & 65.0 & 92.5 \\	
		
	\bottomrule
\end{tabular}
}

\caption{SUS Scores}
\label{tab:sus}
\end{table}

We evaluated each system using the System Usability Scale (SUS).
A breakdown of the SUS score for each system and type of participant (i.e., Participant A---Johnny, Participant B---Jane, or both) is given in Table~\ref{tab:sus}.
The mean value is used as the SUS score \cite{brooke1996sus}.

When evaluating whether a participant's role as Johnny or Jane affected the SUS score, we found a statistically significant difference for Pwm (two-tailed student t-test, equal variance --- $p = 0.05$).
For the other two systems, the differences in SUS scores were slight and were not statistically significant (two-tailed student t-test, equal variance --- Virtru -- $p = 0.67$, Tutanota -- $p = 0.30$).

\begin{figure*}[t]
\centering
\includegraphics[width=0.8\textwidth]{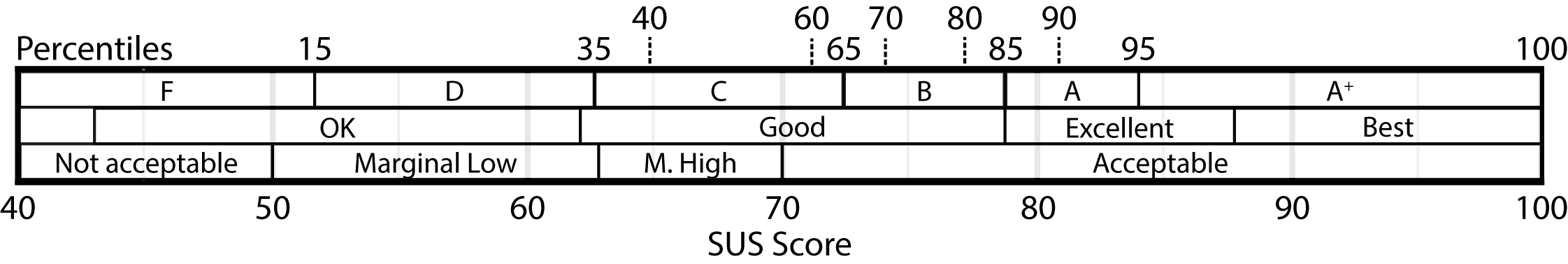}
\caption{Adjective-based Ratings to Help Interpret SUS Scores}
\label{fig:sus}
\end{figure*}

To give greater context to the meaning of each system's SUS score, we leveraged the work of several researchers.
Bangor et al. \cite{bangor2009determining} analyzed 2,324 SUS surveys and derived a set of acceptability ranges that describe whether a system with a given score is acceptable to users in terms of usability.
Bangor et al. also associated specific SUS scores with adjective descriptions of the system's usability.
Using this data, we generated ranges for these adjective ratings, such that a score is correlated with the adjective it is closest to in terms of standard deviations.
Sauro et al. \cite{sauro2011practical} also analyzed SUS scores from Bangor et al. \cite{bangor2008empirical}, Tullis et al. \cite{tullis2004comparison}, and their own data.
They calculated the percentile values for SUS scores and assigned letter grades based on percentile ranges.
These contextual clues are presented in Figure~\ref{fig:sus}.

Pwm and Virtru's SUS scores of 72.7 and 72.3, respectively, are rated as having ``Good'' usability.
Both systems fall right at the \nth{65} percentile and on the line between a ``B'' and ``C'' grade.
The difference between these two systems is not statistically significant (two-tailed student t-test, matched pairs --- $p = 0.86$).
The scores for Pwm are roughly consistent with those seen in prior work~\cite{ruoti2013confused,ruoti2015helping}, though our results exhibited more low outliers than prior studies.
Whether this difference is due to negative experience related to using these systems with a friend as compared to a study coordinator or whether it is due to differences in the study populations is unclear.

Tutanota's score of 52.2 is rated as having ``OK'' usability.
It falls in just about the \nth{15} percentile and at just about the base of the ``D'' grade.
The difference between Tutanota and the other systems (i.e., Pwm and Virtru) is statistically significant (two-tailed student t-test, matched pairs --- $p < 0.001$)



\subsection{Time}

We recorded the time it took each participant to finish the task. Completion times are split into two stages:

\begin{enumerate}
\item Timing for this stage started when Johnny clicked the ``Install'' (Pwm, Virtru) or ``Sign Up'' button. Timing ended when Johnny had successfully sent an encrypted email with their SSN and last year's tax PIN.

\item Timing for this stage started when Jane opened the encrypted email sent in the previous stage. Timing ended after Jane had successfully sent an encrypted email with the confirmation code and this year's tax PIN. This stage included the time Jane spent determining how to decrypt the initial message. In the case of Tutanota, this included obtaining the shared secret from Johnny.

\end{enumerate}

\begin{table}
\begin{center}
\resizebox{\columnwidth}{!}{

\begin{tabular}{l|l|c|cc|ccccc|}

	\rule{0pt}{11ex} & \rot{Stage} & \rot{Count} & \rot{Mean} & \rot{\shortstack[1]{Standard\\Deviation}} & \rot{Min} & \rot{Q1} & \rot{Median} & \rot{Q3} & \rot{Max} \\
	\midrule
	
	Pwm			& 1 	& 25 & 2:29 & 0:49 & 1:32 & 1:46 & 2:14 & 2:45 & 4:09 \\
	Pwm			& 2 	& 24 & 2:59 & 0:52 & 1:24 & 2:23 & 2:51 & 3:27 & 5:05 \\
	Pwm			& Both 	& 49 & 5:28 & 1:22 & 2:56 & 4:38 & 5:08 & 6:17 & 9:14 \\
	\midrule

	Virtru		& 1 	& 25 & 2:39 & 1:09 & 1:23 & 1:51 & 1:51 & 2:54 & 5:46 \\
	Virtru		& 2 	& 23 & 3:06 & 1:53 & 0:59 & 2:02 & 2:50 & 3:22 & 9:22 \\
	Virtru		& Both 	& 48 & 5:48 & 2:55 & 2:27 & 4:02 & 5:07 & 6:12 & 15:08 \\
	\midrule

	Tutanota	& 1 	& 25 & 3:49 & 1:04 & 2:17 & 3:01 & 3:40 & 4:27 & 6:26 \\
	Tutanota	& 2 	& 24 & 5:49 & 3:38 & 1:20 & 3:53 & 5:15 & 6:41 & 18:53 \\
	Tutanota	& Both 	& 49 & 9:41 & 3:54 & 4:11 & 7:45 & 9:06 & 10:49 & 22:26 \\
		
	\bottomrule
\end{tabular}
}
\end{center}

\caption{Time Taken to Complete Task (min:sec)}
\label{tab:time}
\end{table}

\begin{figure}[t]
\centering
\includegraphics[width=1.0\columnwidth]{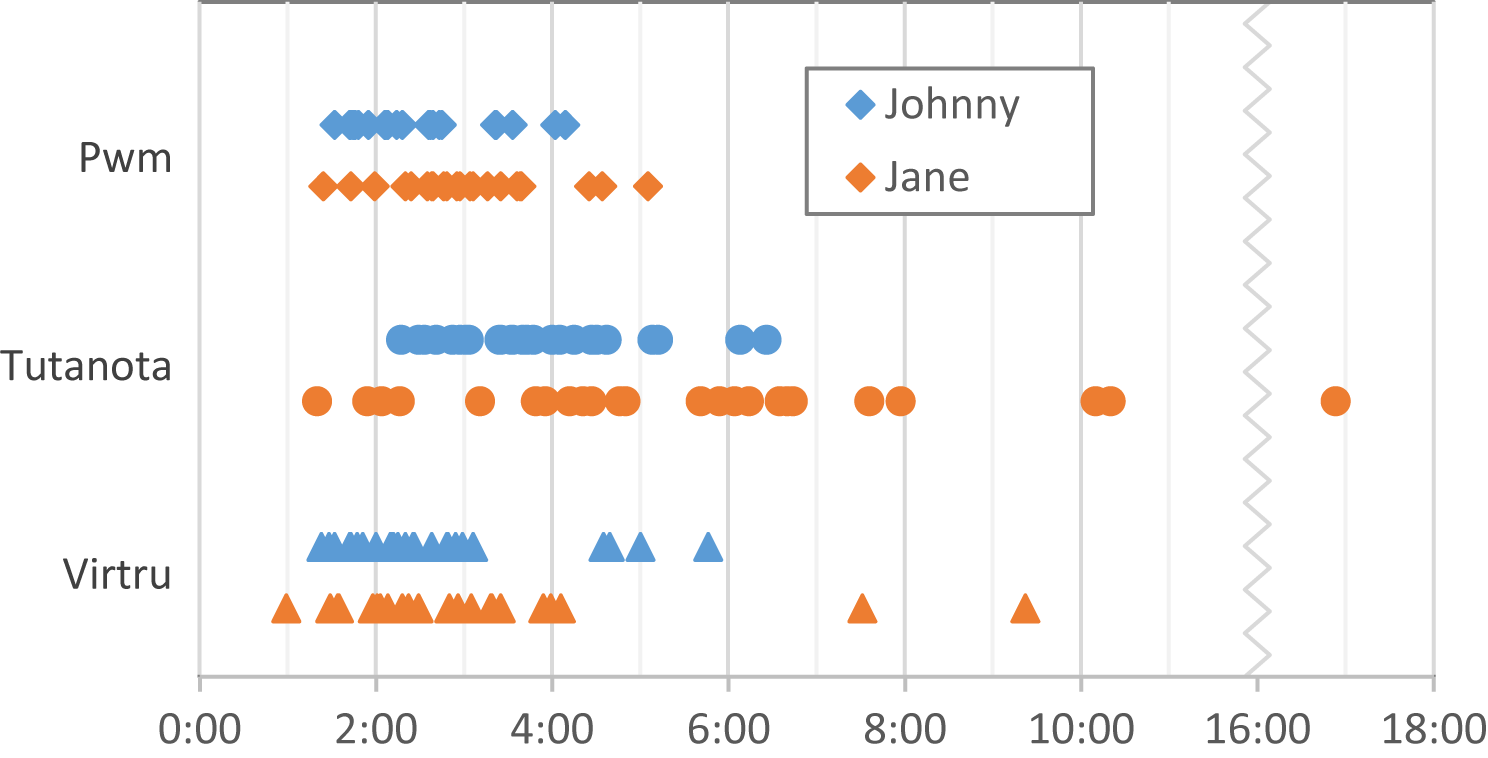}
\caption{Time to complete task stages}
\label{fig:time}
\end{figure}

Timings were calculated using the recorded video.
There were two sessions with abnormalities in the recording.
First, the Virtru portion of a Jane session had become corrupted.
Second, an entire Jane session was also corrupted.
This second session was of special note because this participant failed to successfully complete Stage 2 using Tutanota.
The remaining times are reported in Table~\ref{tab:time} and are graphically shown in Figure~\ref{fig:time}.

In line with the SUS scores, both Pwm and Virtru have completion times that are roughly the same, with any differences failing to be statistically significant (two-tailed student t-test, matched pairs --- Stage 1 -- $p = 0.58$, Stage 2 -- $p = 0.73$, Total -- $p = 0.58$).\footnote{After adjusting the data for the higher number of treatments in which Virtru was the first system tested, we found that on average Johnny's task completion time dropped by 15 seconds for Virtru and increased by 10 seconds for Pwm. These differences are not statistically significant from the non-adjusted data, and the difference between Virtru and Pwm in the adjusted data is not statistically significant.}
Participants using Tutanota took roughly one minute longer to complete Stage 1 and almost three minutes longer to complete Stage 2.
The differences between Tutanota and Pwm/Virtru in Stage 1, Stage 2, and the combined times are all statistically significant (two-tailed student t-test, matched pairs --- each case -- $p < 0.002$)

\subsection{Mistakes}
We defined mistakes as any situation in which sensitive information was sent in plaintext or was sent encrypted along with the key to decrypt the sensitive information (i.e., the Tutanota shared secret was sent as plaintext in email).
Using Pwm, no participants sent their sensitive data in the clear.
With Virtru, only a single participant sent their information in the clear.
In this case, the participant had entered the sensitive information into an unencrypted greeting field that Virtru allows participants to include with an encrypted email.

In contrast, participants were much more likely to make mistakes with Tutanota.
Two-thirds of the participant pairs (68\%, n=17) communicated the Tutanota shared secret over clear text in email.
Additionally, half of the participant pairs (48\%, n=12) selected shared secrets that had low entropy and could be quickly guessed by a password-cracking system.

While Tutanota clearly performed worse than Pwm and Virtru, care should be taken in analyzing this result.
In the post-study interview several participants indicated that while they had transmitted their password over email in the study, they stated that in the real world they would be more likely to send the data over a different channel.

\subsection{Favorite System}
\begin{table}
  \centering
  \small

\begin{tabular}{l|cc|c|}

	\rule{0pt}{1ex} & Johnny & Jane & Total \\
	\midrule
	
	Pwm				& 48\%	& 60\%	& 54\%	\\
	
	Virtru			& 44\%	& 28\%	& 36\%	\\
	
	Tutanota		& 4\%	& 12\%	& 8\%	\\
	\midrule
	
	Disliked All	& 4\%	& 0\%	& 2\%	\\
	\midrule
			
\end{tabular}

\caption{Participants' Favorite Systems}
\label{tab:favorite}
\end{table}

At the end of the study, participants were asked which of the three systems was their favorite.
Their responses are summarized in Table~\ref{tab:favorite}.
Pwm was most frequently rated as the best system, with Virtru also rated highly.
Tutanota was rarely selected as the best system, and one participant indicated that they disliked all of the systems.
These results roughly correlate with the SUS score of each system.

Interestingly, we do see a difference in the choice of favorite system based on what role the participant played.
While Pwm and Virtru are rated as the favorite system about equally by Johnny, Pwm was most often selected as the favorite system by Jane.
Based on participant responses, this disparity is due to the fact that unlike Johnny, Jane had to leave Gmail to interact with Virtru messages, a process that was frequently described negatively.

Similarly, Tutanota was more highly rated by Jane than by Johnny.
Participant responses reveal that this is likely due to the fact that Jane did not have to go through the Tutanota account setup (which required a long, complex password) and selection of a shared secret for the email (which caused nearly all participants to struggle).


\subsection{Differences Based on Treatment}
In Atwater et al.'s work~\cite{atwater2015leading}, they noticed that the order in which users tested systems strongly affected the SUS score for those systems.
We analyzed our results to determine if we saw a similar affect, but found that the order in which the systems were tested had no effect on SUS scores.
This is in line with our prior experience using SUS to evaluate secure email systems~\cite{ruoti2013confused}.

We also evaluated the data to see if the order in which users tested systems affected task completion time, number of mistakes, or participant's favorite systems.
The only measurable difference was in task completion times for Pwm and Virtru: whichever system was tested first would have a longer than average task completion time, and whichever system was tested second would have a lower than average task completion time.
This difference in times was not statistically significant, but this is likely because the sample population was too small.

\section{Discussion}
In this section we discuss themes that we noticed across the study, especially the qualitative feedback provided by participants on the study survey and in the post-study interview.
Participants have all been assigned a unique identifier R[1-25][A,B].
The final letter refers to which role the participant played during the study, and participants with the same number were paired with each other (e.g., R1A and R1B were Johnny and Jane, respectively, in the same study session).

\subsection{Two-Person Studies}
During the study, we noticed several clear benefits of conducting two person studies.

First, by having participants play different roles, we were able to gather data about users' experiences both when they are introduced to secure email and when someone else is introducing them.
For example, in Tutanota, messages need to have a shared secret to be encrypted.
Johnny's experiences revealed the difficulty in discovering that a password is required and needs to be communicated to the recipient, Jane.
Similarly, Jane's experience showed the aversion participants felt to leaving their current email system to view a sensitive message.
While these same experiences might have been elicited by running two different studies, it was convenient to obtain them in a single study, and it was helpful to be able to correlate the experiences of participant pairs.
Furthermore, showing that a participant can successfully use a new secure email system when inducted by another novice user is stronger than only showing that a new user can be inducted by an expert.

Second, this study design led to more natural behaviors by participants.
In past studies, we observed that participants expected study coordinators to immediately respond to emails.
Even after being informed that a response would take a couple of minutes, participants would constantly refresh their inbox to see if a message had arrived, and if a response took longer than fifteen to thirty seconds to arrive participants would often complain.
In contrast, participants in this study were content to wait to receive their email and did not appear agitated when their friends took a long time to respond.
Also, instead of constantly refreshing their inbox, participants would browse the Web or check their phones, which is likely more representative of how they use email in practice.

In addition to these observations by study coordinators, participants also noted that they felt more natural interacting with a friend than with a study coordinator. For example, participants R24B and R25A stated, respectively,

\begin{quote}
\textit{
``...I was more at ease probably than I would've been if it was someone random on the other end...It would've felt more mechanical, robotic, whereas I know [her] and I was calling my wife, `Hi wife! What's the password?' It felt a lot more personable for me I think....''
}
\end{quote}

\begin{quote}
\textit{
``It was good in that you saw the troubles, like the third system [Tutanota], I didn't even know how it worked, so I ended up sending an email to myself on Gmail so then I could see what was happening on her end, to know like how it works on the other end. So I think it's good to have two people on each end that don't know what's going on, because if it weren't I'd assume the person on the other side had done it before...''
}
\end{quote}

Third, participants indicated that because the study was conducted with their friend that they felt more relaxed. For example, participants R11B and R14B indicated, respectively,

\begin{quote}
\textit{
``I thought it was good, I dunno, might've taken the pressure off too, where it's like, `Okay, he's figuring this out too', so I can just, y'know, I don't have to feel as `under-the-microscope' in the study.''
}
\end{quote}

\begin{quote}
\textit{
``I felt like neither of us knew what we were doing, but if I knew that someone else knew what was going on, I'd be like, `K, hopefully I'm not doing it wrong,' so it's kind of like, `K, we're on the same page, neither of us know what we're doing.'\thinspace''
}
\end{quote}


Fourth, we were pleased to note that requiring participants to bring a friend with them resulted in a much lower missed-appointment rate than we have seen in the past.

Based on our observation of participants' behavior and the participants' qualitative feedback, we feel that there is significant value in conducting two person studies.
Still, future research should examine in greater depth the differences between one and two person studies.
For example, an A/B study comparing these two methodologies could be conducted which compares difference in system metrics (e.g., SUS, task completion time) as well measures difference in users agitation during the study (e.g., heart rate, eye tracking).
Similarly, research could compare how participant experiences differ when both roles are filled by a novice, as opposed to having one simulated by a coordinator.

\subsection{Hidden Details and Trust}
Providing further evidence to prior work \cite{fahl2012helping,ruoti2013confused}, participants' experiences demonstrated that when security details were hidden from them, they were less likely to trust the system.
This was most clearly demonstrated when examining Pwm and Virtru.
Both systems use email-based identification and authentication \cite{garfinkel2003email} to verify the user's identity to a key escrow server (i.e., it sends an email to the user with a link to click on to verify their identity).
The difference is that while Virtru requires users to manually open this email and click on the link, Pwm performs this task automatically for users.
While this difference might seem small, it was cause for concern for several participants.
For example, participants R6B and R10B expressed, respectively,

\begin{quote}
\textit{
``I liked the way that one [Virtru] and the last one [Tutanota] both had ways to confirm that it was you and no one else could see the information.''
}
\end{quote}

\begin{quote}
\textit{
``(Interviewer: But you didn't think that [Pwm] was as secure [Virtru]?) [Pwm] \textbf{said} that it was, but I liked how the other ones had additional 'send-you-an-email' verification or a password between you and the other person in the email. Just an added measure to feel like there really is something different. 'Cause Pwm for all I know, like, I'm just taking their word for it. There's not really anything extra that shows that it really is secure.''
}
\end{quote}

In contrast, the shared secret used by Tutanota made it clear that only the recipient who had the password would be able to read the message.
This made a large number of participants feel that Tutanota was the most secure system, even if usability issues prevented it from being their favorite system.
R17B's response demonstrate this principle:

\begin{quote}
\textit{
``Like the order of the programs was interesting 'cause I thought the Virtru one was great, like until I saw this [Tutanota], `Oh, this one [Tutanota] requires a password - why did I think that one [Virtru] was great?' And I wish, it \textbf{would} have required a password because anybody that has your email password can just see [everything]."
}
\end{quote}

The sentiment regarding passwords was so strong that several participants stated that they wished Pwm and Virtru would also allow them to password-encrypt messages.
For example, R17B, R10A, and R10B expressed, respectively, 

\begin{quote}
\textit{
``I like that [Pwm] encrypted the info so that Gmail couldn't read it. I think Pwm would be the best one if it required passwords.''
}
\end{quote}

\begin{quote}
\textit{
``R10B: I would say exactly the second one [Pwm], just with a password (R10A: 'Yeah') per conversation is what I would do. Just because it's so simple, right there. R10A: Stay on Gmail, but then have a password to get to \textbf{that} encrypted email. (R10B: 'Yeah')''
}
\end{quote}

Still, not all participants were enamored with using a shared password, seeing it as an added memory burden or hassle.
As stated by participants R10A, R5A, and R25A, respectively,

\begin{quote}
\textit{
``I will never remember my crazy password.''
}
\end{quote}

\begin{quote}
\textit{
``I don't know if I loved the password idea, just because if I was sending a secure password over something, then why didn't I just send the information over that anyways?''
}
\end{quote}

\begin{quote}
\textit{
``How do you send a password safely if your encrypting program requires a password?''
}
\end{quote}

Some participants were concerned that it was impossible to verify if any of the systems were truly encrypting their data.
This likely stems from two facts: first, that participants are not security experts and lack the means to truly verify the security of a tool, and second, that the tools themselves---once working---never show the user any indication that they are actually receiving encrypted email.
While results from Atwater et al. suggest that showing ciphertext does not address this issue \cite{atwater2015leading}, the fact that participants are concerned indicates that this problem needs more research.
For example, participants R14A 
and R17B stated, respectively,

\begin{quote}
\textit{
``It would be kind of cool to see what it would look like as an encrypted message. ...Seems kind of weird. Like `it's encrypted now, trust us.'\thinspace''
}
\end{quote}


\begin{quote}
\textit{
``I would like to know exactly how the encryption happens - I understand that it is encrypting it, but how do I know it's completely safe?...There are too many programs that are not what they seem, and I would not want this to be one of those.''
}
\end{quote}


\subsection{Integrated vs Depot}
Participants overwhelmingly preferred secure email to be integrated into their existing email systems and not require a second account (i.e., a depot).
This preference was expressed through the low SUS scores of Tutanota and the fact that only four participants rated it as their favorite system.
Additionally, participant comments made it clear that they were not interested in using depot-based secure email.
For example, participants 
R25A and R16A stated, respectively,


\begin{quote}
\textit{
``No one wants another email system.''
}
\end{quote}

\begin{quote}
\textit{
``It is just not my type. I don't want to set up another account and send a password to my friend.''
}
\end{quote}

However, several participants felt that Tutanota was more secure than other systems, precisely because it required the creation of an account separate from Gmail.
Participants noted that in Pwm and Virtru, access to the user's Gmail account was all that was required to decrypt sensitive email.
For example, participants R4B, R25A, and R25B stated, respectively,

\begin{quote}
\textit{
``[I like it,] I dunno, just because I leave my email up a lot, someone could just go on to my email and look at it. I don't sign out of my email.''
}
\end{quote}



\begin{quote}
\textit{
``I just kinda feel like anybody could go into your email and look at those secure ones if it \textbf{is} inside your email...''
}
\end{quote}

\begin{quote}
\textit{
``How strong is your Gmail password, you know? If you can get in there, then it defeats these other encryption. So, really, you're just trying to hide your stuff from Google, which, they already know everything, so.''
}
\end{quote}


\subsection{Tutorials}
Tutorials were a significant factor in participants' experiences.
Pwm was rated by participants as having the best tutorials, with a fourth of participants (24\%, n=12) bringing up tutorials when asked what they liked about Pwm.
Participants largely liked the style of the tutorials as well as their content. For example, participant 8B expressed, \textit{``I also really liked the tutorial. It was similar to tutorials Apple or Google/Gmail give you to learn things.''}

Virtru also has tutorials, but praise for these tutorials was not as common as it was with Pwm, with Jane participants criticizing the tutorials more than Johnny participants.
This result can likely be attributed to the fact that the Virtru plugin walks new users through a tutorial upon installation, but someone who receives a Virtru-encrypted message without the plugin is simply presented with a blue button labeled "Unlock message" without additional instruction beyond what the sender of the email has personally and manually added. 
This is in contrast to Pwm, which prefaces incoming encrypted email with instructions on what encrypted email is and how the recipient should go about decrypting the message.

Tutanota had no tutorials, and this clearly led to confusion.
Nearly all participants failed to notice that they needed to set a password to encrypt their email, and just as many didn't realize that they needed to communicate this password to the other participant.
Additionally, some participants didn't understand that they couldn't just use Tutanota to communicate the password
Many of these problems could have been alleviated by a simple tutorial.

\subsection{Reasons to Use Encrypted Email}
The majority of participants (72\%) agreed with the survey statement, ``I want to be able to encrypt my email,'' although only a much smaller fraction (20\%) agreed with the idea that they would ``encrypt email frequently.''
Still, when asked to describe how they would use encrypted email in practice, many participants were unsure.
The range of opinions are summarized in responses from participants R22A, R24A, R20A, and R23B, respectively, on how they would use secure email in practice:

\begin{quote}
\textit{
``Um, I've never really used it before because I didn't know it was so accessible through Gmail so now that I know I can, I will use it more often. (Interviewer: Do you actually plan to use it?) Yeah, I will.''
}
\end{quote}

\begin{quote}
\textit{
``Like, just the other day I needed [my husband's] social security card number for something and then didn't feel like there was any way I could ask him and if I had known about this, I would have done that.''
}
\end{quote}


\begin{quote}
\textit{
``Well, I'm trying to think when I would need to. It would be nice to have it, in case, but I don't know if there's anyone I would need to send that information to.''
}
\end{quote}

\begin{quote}
\textit{
``Knowing that I could encrypt email I probably could find uses for it, but...''
}
\end{quote}

These responses indicate that more research needs to be done to discover under what circumstances the users would employ secure email.

\section{Conclusion}

In this work, we conducted the first two-person study of secure email where two novice users are brought into the lab together and asked to exchange secure email between themselves. Our study analyzed Pwm, Tutanota, and Virtru.
Using a two-person study enabled us see participants under different first-use experiences. In addition,
participants exhibited more natural behaviors, seemed less agitated, and indicated that they felt less like they were ``under the microscope.''

Our results indicate several observations about secure email systems.
First, we found further evidence that hiding the security details can lead to a lack of trust in the secure email system. This gives further credence to similar results from earlier work~\cite{ruoti2013confused}.
Second, we found that participants largely rejected depot-based secure email systems.
Third, participant success in using a system without mistakes is heavily influenced by the presence of well-designed tutorials.
Lastly, while participants are interested in using secure email, few express a desire to use it regularly and most are unsure of when or how they would use it in practice.

\balance
\bibliographystyle{SIGCHI-Reference-Format}
\bibliography{paper}

\end{document}